\documentclass{article}

\usepackage{arxivsimplified}

\usepackage[utf8]{inputenc} 
\usepackage[T1]{fontenc}    
\usepackage{hyperref}       
\usepackage{url}            
\usepackage{booktabs}       
\usepackage{amsfonts}       
\usepackage{nicefrac}       
\usepackage{microtype}      
\usepackage{lipsum}

\usepackage{graphicx}
\graphicspath{{figs_png/}}
\usepackage{subfigure}
\usepackage{cite}
\usepackage[dvipsnames]{xcolor}
\usepackage{amsmath}
\DeclareMathAlphabet\mathbfcal{OMS}{cmsy}{b}{n}

\usepackage{blindtext}
\usepackage{amsfonts}
\usepackage{multirow}
\usepackage{array}
\usepackage{mathtools}
\usepackage{listings}
\usepackage{xcolor}
\definecolor{codegreen}{rgb}{0,0.6,0}
\definecolor{codegray}{rgb}{0.5,0.5,0.5}
\definecolor{codepurple}{rgb}{0.58,0,0.82}
\definecolor{backcolour}{rgb}{0.95,0.95,0.92}
\usepackage{color}
\usepackage{xcolor}
\definecolor{rev1}{rgb}{0,0,0}
\usepackage{esvect}
\usepackage{tabularx}
\newcolumntype{L}[1]{>{\raggedright\arraybackslash}p{#1}}
\newcolumntype{C}[1]{>{\centering\arraybackslash}p{#1}}
\newcolumntype{R}[1]{>{\raggedleft\arraybackslash}p{#1}}

\usepackage{algorithm}
\usepackage{algorithmic}
\usepackage{enumitem}
\setlist[itemize]{leftmargin=*}

\lstdefinestyle{mystyle}{
    backgroundcolor=\color{backcolour},   
    commentstyle=\color{codegreen},
    keywordstyle=\color{magenta},
    numberstyle=\tiny\color{codegray},
    stringstyle=\color{codepurple},
    basicstyle=\ttfamily\footnotesize,
    breakatwhitespace=false,         
    breaklines=true,                 
    captionpos=b,                    
    keepspaces=true,                 
    numbers=left,                    
    numbersep=5pt,                  
    showspaces=false,                
    showstringspaces=false,
    showtabs=false,                  
    tabsize=2
}
 
\usepackage{scalerel,stackengine}
\stackMath
\newcommand\reallywidehat[1]{%
\savestack{\tmpbox}{\stretchto{%
  \scaleto{%
    \scalerel*[\widthof{\ensuremath{#1}}]{\kern-.6pt\bigwedge\kern-.6pt}%
    {\rule[-\textheight/2]{1ex}{\textheight}}
  }{\textheight}%
}{0.5ex}}%
\stackon[1pt]{#1}{\tmpbox}%
}

\title{Equation-free surrogate modeling of geophysical flows at the intersection of machine learning and data assimilation}

\author{
    Suraj Pawar  \\
    School of Mechanical \& Aerospace Engineering,\\
    Oklahoma State University, \\
    Stillwater, Oklahoma - 74078, USA.\\
    \texttt{supawar@okstate.edu} \\
    \And
    Omer San \\
    School of Mechanical \& Aerospace Engineering,\\
    Oklahoma State University, \\
    Stillwater, Oklahoma - 74078, USA.\\
    \texttt{osan@okstate.edu} \\
}

\begin{document}
\maketitle

\begin{abstract}
There is a growing interest in developing data-driven reduced-order models for atmospheric and oceanic flows that are trained on data obtained either from high-resolution simulations or satellite observations. The data-driven models are non-intrusive in nature and offer significant computational savings compared to large-scale numerical models. These low-dimensional models can be utilized to reduce the computational burden of generating forecasts and estimating model uncertainty without losing the key information needed for data assimilation to produce accurate state estimates. 
This paper aims at exploring an equation-free surrogate modeling approach at the intersection of machine learning and data assimilation in Earth system modeling. With this objective, we introduce an end-to-end non-intrusive reduced-order modeling (NIROM) framework equipped with contributions in modal decomposition, time series prediction, optimal sensor placement, and sequential data assimilation. Specifically, we use proper orthogonal decomposition (POD) to identify the dominant structures of the flow, and a long short-term memory network to model the dynamics of the POD modes. The NIROM is integrated within the deterministic ensemble Kalman filter (DEnKF) to incorporate sparse and noisy observations at optimal sensor locations obtained through QR pivoting. The feasibility and the benefit of the proposed framework are demonstrated for the NOAA Optimum Interpolation Sea Surface Temperature (SST) V2 dataset. Our results indicate that the NIROM is stable for long-term forecasting and can model dynamics of SST with a reasonable level of accuracy. Furthermore, the prediction accuracy of the NIROM gets improved by almost one order of magnitude by the DEnKF algorithm. This work provides a way forward toward transitioning these methods to fuse information from large-scale Earth system models and diverse observations to achieve accurate forecasts.   
\end{abstract}


\section{Introduction}

The integration of models and observations has greatly transformed the predictions of the Earth system, including weather forecasts and climate projections \cite{gettelman2022future, bauer2021digital}. Observations can be either used to estimate the initial condition for the prediction that is consistent with the present state of the Earth system through the process of data assimilation or to reduce model errors by advancing the representation of certain processes within a model. Data assimilation (DA) is a well-established method that involves combining information coming from the forecast model with available observations and are used extensively in numerical weather prediction (NWP) \cite{lorenc2003potential,rabier2005overview, Navon2009}. DA can also be viewed from the Bayesian perspective that involves fusing data (observations) with the prior knowledge (i.e., mathematical representation of physical processes; model output) to obtain an estimate of the distribution of the true state of the process \cite{wikle2007bayesian, kalnay2003atmospheric}. 

DA methods are usually classified into two types, variational approaches and sequential approaches \cite{lewis2006dynamic}. Variational DA is formulated as the constrained optimization problem defined by a cost function to minimize the discrepancy between the prior knowledge (i.e., the computational model) and observations. On the other hand, the sequential DA involves evolving the state of the system with background information until observations get available, and then updating the system's state. One of the key components of the DA cycle is the forecast model. The forecast models used within DA are based on solving the governing equations or the best approximation of physical processes numerically using spatial and temporal discretization on a computational grid \cite{bauer2018quiet}. Despite their success, the current forecast models have difficulty representing complex processes like turbulence, convection and clouds \cite{yano2018scientific, randall2018years} leading to high uncertainty in the prediction \cite{zelinka2020causes}. Additionally, the computational cost of ensemble forecasting (using many realizations with perturbed initial conditions) with numerical models is huge. As a result, the forecast model is one of the major limiting factors in DA, and this has spurred interest in using data-driven methods for the Earth system modeling that can deliver both computational efficiency and better representation of physical processes derived from data \cite{dueben2018challenges,boukabara2021outlook}.  

Data-driven methods have been applied to a wide range of problems in Earth system modeling \cite{reichstein2019deep}. For example, machine learning approaches have been explored for parameterization of subgrid-scale processes \cite{rasp2018deep, gagne2020machine, bolton2019applications}, precipitation nowcasting \cite{shi2015convolutional}, superresolution of wind and solar data \cite{stengel2020adversarial}, and weather forecasting \cite{weyn2019can}. Some of the studies with the data-driven weather prediction (DDWP) have already started showing promising results with similar if not superior performance compared to state-of-the-art NWP models \cite{schultz2021can, pathak2022fourcastnet, rasp2021data}. These DDWP models leverage deep learning methods like a convolutional neural network, Fourier neural operator, and recurrent neural network that are trained on data obtained from reanalysis products \cite{rasp2020weatherbench}. Furthermore, data-driven models can be augmented by taking up the information from physical knowledge, or at least respecting the conservation properties \cite{kashinath2021physics, karniadakis2021physics, chantry2021opportunities, irrgang2021towards}. Some of the work towards these directions include enforcing the physical laws or statistical constraint into the loss function \cite{raissi2019physics, wu2020enforcing}, tailoring the architecture design to enforce certain symmetry or conservation laws \cite{mohan2020embedding, beucler2021enforcing, pawar2022frame}, symbolic regression for equation discovery \cite{zanna2020data}, and end-to-end learning strategy \cite{frezat2022posteriori}. 

These studies indicate that ML has the potential to improve scientific knowledge (and hence models), especially when we cannot express our understanding in the form of mathematical equations in physics-based models \cite{geer2021learning,buizza2022data}. DA could benefit a lot from ML, where a forecast model is replaced with a hybrid model that incorporates a neural network as a component of the physical model \cite{hsieh1998applying, pawar2021data} or as a complete replacement with a data-driven model for forecasting and state estimation \cite{chattopadhyay2022towards, penny2022integrating}. Another advantage of an ML-based emulator is the quick computation of backpropagation gradient with automatic differentiation that can be used as a replacement for expensive adjoint solvers within variational DA \cite{maulik2022aieada, chennault2021adjoint}. Similarly, some of the challenges with ML such as handling uncertain and sparsely sampled data can be mitigated with DA. Brajard et al. \cite{brajard2021combining} proposed a two-step process to learn the parameterization of unresolved processes where a neural network was trained to learn the model error obtained from the analysis state of the truncated model. Some other works at the intersection of DA and ML include iterative application of neural network and DA to learn the chaotic dynamics \cite{brajard2020combining}, enforcing conservation of mass constraint in DA \cite{ruckstuhl2021training}, and for building tangent-linear and adjoint models of parameterization schemes in variational DA \cite{hatfield2021building}.   

In this work, we seek to build a data-driven surrogate model trained on the analysis data and integrate it with a sequential DA cycle. More specifically, we will focus on a class of data-driven methods that are based on projection-based reduced-order modeling (ROM). These methods first extract the recurrent spatial structures from the high-dimensional data using singular value decomposition or convolutional autoencoder \cite{brunton2020machine, mou2020reduced}. The next step is to either solve the projected governing equations on the dominant modes \cite{san2015stabilized, carlberg2011efficient} or learn the projected dynamics with data-driven methods \cite{wang2018model, rahman2019nonintrusive, pan2018long}. The recurrent neural network is one of the most popular algorithms in modeling the dynamics of lower-dimensional latent space, and has been applied for a wide range of applications \cite{yu2019non,ahmed2021closures}. One of the advantages of using projection-based ROM as a surrogate model is that it allows for DA on a reduced-order space rather than on a full-state and this latent assimilation framework leads to substantial speed improvement in contrast to DA \cite{quilodran2018fast, amendola2020data, liu2022enkf, peyron2021latent}. The key challenge with latent assimilation is to link the real-time observations in physical space with the latent observation space \cite{cheng2022generalised}. Our approach for latent observation space construction is by first identifying the near-optimal sensor locations using QR pivoting and then using the measurements at these discrete locations for reconstruction on a tailored basis \cite{manohar2018data}.  

The paper is organized as follows. In Section~\ref{sec:surrogate}, the surrogate modeling with proper orthogonal decomposition (POD) for dimensionality reduction, and a long short-term memory (LSTM) network for modeling dynamics is discussed. The detailed discussion on the deterministic ensemble Kalman filter (DEnKF) is provided in Section~\ref{sec:da}. The computation of the latent observation space is described in Section~\ref{sec:qr}. In Section~\ref{sec:results}, the performance of the proposed framework is demonstrated for the NOAA Optimum Interpolation (OI) Sea Surface Temperature (SST) V2 dataset. Finally, concluding remarks are provided in Section~\ref{sec:conlusion}.


\section{Methods} \label{sec:methods}
In this section, we introduce different components of the proposed framework including surrogate modeling, sequential data assimilation, and optimal sensor placement strategy. The surrogate model is based on the projection-based model order reduction which relies on proper orthogonal decomposition (POD) and a recurrent neural network to model the dynamics of the latent space. This surrogate model is integrated within the sequential data assimilation to correct the forecast using real-time observations. The latent observation space is obtained by first reconstructing the full state from point sensor measurements and then projecting it onto POD bases.   

\subsection{Surrogate Modeling} \label{sec:surrogate}
Our surrogate model is constructed using the projection-based reduced-order modeling where the high-dimensional system is first compressed to a low-dimensional system and the dynamics of the low-dimensional system is modeled. The compression is achieved by projecting the high-dimensional data onto a set of optimal linear basis functions obtained via POD \cite{sirovich1987turbulence,berkooz1993proper}. POD provides the optimal linear basis functions as they minimize the error between the true data and its projection in the $L_2$ sense compared to all other linear basis functions of the same dimension. Given the $N_s$ snapshots of the data for a state variable $\boldsymbol{x} \in \mathbb{R}^n$, we can form the matrix $\boldsymbol{A}$ as follows
\begin{equation}
    \boldsymbol{A} = \bigg[ \boldsymbol{x}^{(1)},\boldsymbol{x}^{(2)}, \dots, \boldsymbol{x}^{(N_s)}\bigg] \in \mathbb{R}^{n \times N_s}, \label{eq:snapshot_A}
\end{equation}
where $\boldsymbol{x}^{(k)}$ corresponds to an individual snapshot in time. Then, we perform the reduced singular value decomposition (SVD) as follows
\begin{equation}
    \boldsymbol{A} = \boldsymbol{U}\boldsymbol{\Sigma}\boldsymbol{V}^T = \sum_{k=1}^{N_s}\sigma_k \boldsymbol{u}_k \boldsymbol{v}_k^T,
\end{equation}
where $\boldsymbol{U}\in \mathbb{R}^{n \times N_s}$ is a matrix with orthonormal columns which corresponds to the left-singular vectors, $\boldsymbol{V} \in \mathbb{R}^{N_s \times N_s}$ is matrix with orthonormal columns representing the right-singular vectors, and $\boldsymbol{\Sigma} \in \mathbb{R}^{N_s \times N_s}$ is a matrix with non-negative diagonal entries, called singular values, and are arranged such that $\sigma_1 \ge \sigma_2 \ge \dots \ge \sigma_{N_s} \ge 0$. For the dimensionality reduction task, only the first $N_r$ columns of $\boldsymbol{U}$ and $\boldsymbol{V}$ (denoted as $\overline{\boldsymbol{U}}$ and $\overline{\boldsymbol{V}}$) are retained along with the upper-left $N_r \times N_r$ sub-matrix of $\boldsymbol{\Sigma}$ (denoted as $\overline{\boldsymbol{\Sigma}}$). The reduced-order approximation of the matrix $\boldsymbol{A}$ can be written as follows
\begin{equation}
    \boldsymbol{\overline{A}} = \boldsymbol{\overline{U}}~\boldsymbol{\overline{\Sigma}}~\boldsymbol{\overline{V}}^T.
\end{equation}

The total $L_2$ error between the snapshot data matrix $\boldsymbol{A}$ and its reduced-order approximation $\boldsymbol{\overline{A}}$ satisfies the following equalities \cite{trefethen1997numerical} 
\begin{equation}
    \lVert \boldsymbol{A} - \boldsymbol{\overline{A}} \rVert_2 ~=~ \underset{\text{rank}(\boldsymbol{B}) \le N_r}{\underset{\boldsymbol{B} \in \mathbb{R}^{n \times N_s}}{\inf}} ~\lVert \boldsymbol{A} - \boldsymbol{B} \rVert_2 ~=~ \sigma_{N_r+1},\label{eq:error_bound}
\end{equation}
where $\lVert \cdot \rVert_2$ is the matrix-2 norm. Equation~\ref{eq:error_bound} states that across all possible matrices $\boldsymbol{B} \in \mathbb{R}^{n \times N_s}$ of rank $N_r$ or less, the matrix $\boldsymbol{\overline{A}}$ provides the best approximation in the $L_2$ sense and the error between $\boldsymbol{A}$  and its $N_r$-order approximation $\boldsymbol{\overline{A}}$ equals $\sigma_{N_r+1}$. From here on, the $N_r$ columns of $\boldsymbol{U}$ are called as the POD modes or basis functions and we denote them as $\boldsymbol{\Phi} = \{ \phi_k\}_{k=1}^{N_r}$. Once the POD modes are obtained, the compressed latent space for a single state vector $\boldsymbol{x}$ can be written as 
\begin{equation}
    {\boldsymbol{\alpha}} = \boldsymbol{\Phi}^T \boldsymbol{x}, \label{eq:pod_coeffients}
\end{equation}
where ${\boldsymbol{\alpha}}$ is the reduced-order approximation of the full state vector $\boldsymbol{x}$ and is also referred to as the POD modal coefficients. The reconstruction of the state vector is then computed as
\begin{equation}
    \overline{\boldsymbol{x}} ~=~ \boldsymbol{\Phi}{\boldsymbol{\alpha}} ~=~ \boldsymbol{\Phi}\boldsymbol{\Phi}^T \boldsymbol{x}, \label{eq:reconstruction}
\end{equation}
where $\overline{\boldsymbol{x}}$ is the optimal reconstruction of full state vector $\boldsymbol{x}$. The number of retained modes, i.e., $N_r$, is decided based on the variance of the data captured by the retained modes. The singular values ${\sigma_i}$ give a measure of the quality of information that is retailed in $N_r$-order approximation of the matrix $\boldsymbol{A}$. The amount of the energy retained by POD modes can be calculated using the quantity called the relative information content (RIC) as follows
\begin{equation}\label{eq:ric}
    \text{RIC}(N_r) = \frac{\sum_{j=1}^{N_r} \sigma_j^2}{\sum_{j=1}^{N_s} \sigma_j^2}    
\end{equation}

The second step of the surrogate modeling is to model the evolution of the latent variables. If the exact equations governing the physical system are known, intrusive approaches like Galerkin projection can be applied to build the ROM. The ROM is obtained by substituting the low-rank approximation into the full order model and then taking the inner product with test basis functions to yield a system of $N_r$ ordinary differential equations (ODEs). However, for geophysical systems, the exact governing equations are unavailable or insufficient for the desired purpose (for example due to coarse grid resolution). On the other hand, the reasonably accurate and complete observational data for the evolution of the state of the system has been collected for many decades. This situation makes the equation-free techniques for predicting the future state of the dynamical system very attractive for multiscale and chaotic dynamical system modeling \cite{kevrekidis2009equation, vlachas2018data, pathak2018model}. Recently, machine learning methods based on recurrent neural network have been applied successfully in many studies to build the non-intrusive ROM \cite{yu2019non}.

We apply the long short-term memory (LSTM) neural network to model the evolution of the latent variables. LSTM has been successfully applied in many studies dealing with modeling high-dimensional spatio-temporal chaotic dynamics of physical systems \cite{rahman2019nonintrusive,mohan2018deep,vlachas2020backpropagation}. LSTM is a type of recurrent neural network (RNN) that can capture the long-term dependencies in the evolution of time series data \cite{hochreiter1997long}. RNNs contain loops that allow them to persist information from one time step to another and can be expressed as
\begin{align}
    \mathbf{h}^{(t)} &= f_{h \rightarrow h} (\mathbf{o}^{(t)}, \mathbf{h}^{(t-1)}), \\
    \tilde{\mathbf{o}}^{(t+1)} &= f_{h \rightarrow o} (\mathbf{h}^{(t)}),
\end{align}
where $\mathbf{h}^{(t)} \in \mathbb{R}^{d_h}$ is the hidden state at time $t$, $\mathbf{o}^{(t)} \in \mathbb{R}^{d_h}$ is the input vector at time $t$, $f_{h \rightarrow h}$ is the hidden to hidden mapping, and $f_{h \rightarrow o}$ is the hidden to output mapping. The output of the model is the forecast $\tilde{\mathbf{o}}^{(t+1)}$ at time step $t+1$. 

One of the limitations of RNNs is vanishing (or exploding) gradient to capture the long-term dependencies. This problem occurs because the gradient is multiplied with the weight matrix repetitively during backpropagation through time (BPTT) \cite{werbos1990backpropagation}. The LSTM mitigates this issue by employing a memory cell composed of gating mechanism that decides which information to memorized or forgotten. The equations that implicitly define the mapping from hidden state of the previous time step (i.e., $\mathbf{h}^{(t-1)}$) and input vector at the current time step (i.e., $\mathbf{o}^{(t)}$) to the forecast hidden state (i.e., $\mathbf{h}^{(t)}$) can be written as follows
\begin{align}
    \mathbf{g}^{(t)}_f &= \sigma(\mathbf{W}_f [\mathbf{h}^{(t-1)},\mathbf{o}^{(t)}] + \mathbf{b}_f), \\
    \mathbf{g}^{(t)}_i &= \sigma(\mathbf{W}_i [\mathbf{h}^{(t-1)},\mathbf{o}^{(t)}] + \mathbf{b}_i), \\
    \tilde{\mathbf{c}}^{(t)} &= \text{tanh}(\mathbf{W}_c [\mathbf{h}^{(t-1)},\mathbf{o}^{(t)}] + \mathbf{b}_c), \\
    \mathbf{c}^{(t)} &= \mathbf{g}^{(t)}_f \odot \mathbf{c}^{(t-1)} + \mathbf{g}^{(t)}_i \odot \tilde{\mathbf{c}}^{(t)}, \\
    \mathbf{g}^{(t)}_o &= \sigma(\mathbf{W}_o [\mathbf{h}^{(t-1)},\mathbf{o}^{(t)}] + \mathbf{b}_o), \\
    \mathbf{h}^{(t)} &= \mathbf{g}^{(t)}_o \odot \text{tanh} (\mathbf{c}^{(t)}),
\end{align}
where $\mathbf{g}^{(t)}_f,~\mathbf{g}^{(t)}_i,~\mathbf{g}^{(t)}_o \in \mathbb{R}^{d_h}$ are the forget gate, input gate, and output gate, respectively. The $\mathbf{o}^{(t)} \in \mathbb{R}^{d_i}$ is the input vector at time $t$, $\mathbf{h}^{(t)} \in \mathbb{R}^{d_h}$ is the hidden state, $\mathbf{c}^{(t)} \in \mathbb{R}^{d_h}$ is the cell state, $\mathbf{W}_f,~\mathbf{W}_i,~\mathbf{W}_c,~\mathbf{W}_o \in \mathbb{R}^{d_h \times (d_h + d_i)}$ are the weight matrices, and $\mathbf{b}_f,~\mathbf{b}_i,~\mathbf{b}_c,~\mathbf{b}_o \in \mathbb{R}^{d_h}$ are the bias vectors. The symbol $\odot$ denotes the element-wise multiplication, and $\sigma$ is the sigmoid activation function, i.e., $\sigma(x)=(1/(1+e^{-x}))$. The above set of equations are unfolded in time to model the temporal dependencies in predicting future state $\mathbf{o}^{(t+1)}$ given $\mathbf{o}^{(t)},\mathbf{o}^{(t-1)},\cdots,\mathbf{o}^{(t-l)}$. The $l$ is referred to as the lookback window which governs how much amount of the old temporal information is required to forecast the future state of the system accurately. 


When we utilize the LSTM network for constructing a surrogate model, the reduced-order state of the system at a future time step, i.e., ${\boldsymbol{\alpha}}^{(k+1)}$, is learned as the function of a short history of $l$ past temporally consecutive reduced-order states as follows 
\begin{equation}
    {\boldsymbol{\alpha}}^{(k+1)} = \boldsymbol{F}(\underbrace{{\boldsymbol{\alpha}}^{(k)}, {\boldsymbol{\alpha}}^{(k-1)}, \dots , {\boldsymbol{\alpha}}^{(k-l+1)}}_{{\boldsymbol{\alpha}}^{(k):(k-l+1)}}; \boldsymbol{\theta}),
\end{equation}
where $\boldsymbol{F}(\cdot ~ ; \boldsymbol{\theta})$ is the nonlinear function parameterized by a set of parameters $\boldsymbol{\theta}$, and ${\boldsymbol{\alpha}}$ is the low-dimensional approximation of the full state vector, i.e., the POD modal coefficients given in Eq.~\ref{eq:pod_coeffients}. Since the surrogate model does not use any governing equations of the system, it is also referred to as the non-intrusive reduced-order model (NIROM).

\subsection{Data Assimilation} \label{sec:da}
We consider the dynamical system whose evolution can be represented as 
\begin{equation}\label{eq:dyn_model}
    \boldsymbol{x}^{(k+1)} = \boldsymbol{M}_{t_k \rightarrow t_{k+1}}(\boldsymbol{x}^{(k)}) + \boldsymbol{w}^{(k+1)},
\end{equation}
where $\boldsymbol{x}^{(k)} \in \mathbb{R}^n$ is the state of the system at discrete time $t_k$, and $\boldsymbol{M}_{t_k \rightarrow t_{k+1}}:\mathbb{R}^n \rightarrow \mathbb{R}^n$ is the nonlinear model operator that defines the evolution of the system over the interval $[t_k,t_{k+1}]$. The term $\boldsymbol{w}^{(k+1)}$ denotes the model error that takes into account any type of uncertainty in the model that can be attributed to boundary conditions, imperfect models, etc. Let $\boldsymbol{z}^{(k)} \in \mathbb{R}^m$ be observations of the state vector obtained from sparse, noisy measurements and can be written as
\begin{equation}
    \boldsymbol{z}^{(k+1)} = \boldsymbol{q}(\boldsymbol{x}^{(k+1)}) + \boldsymbol{v}^{(k+1)},
\end{equation}
where $\boldsymbol{q}(\cdot)$ is a nonlinear function that maps $\mathbb{R}^n \rightarrow \mathbb{R}^m$, and $\boldsymbol{v}^{(k+1)} \in \mathbb{R}^m$ is the measurement noise. We assume that the measurement noise is a white Gaussian noise with zero mean and the covariance matrix $\boldsymbol{R}^{(k+1)}$, i.e., $\boldsymbol{v}^{(k+1)} \sim {\cal{N}}(0,\boldsymbol{R}^{(k+1)})$. Additionally, the noise vectors $\boldsymbol{w}^{(k+1)}$ and $\boldsymbol{v}^{(k+1)}$ are assumed to be uncorrelated to each other at all time steps. 

The sequential DA can be considered as a Bayesian inference framework that estimates the state $\boldsymbol{x}^{(k+1)}$ of the system given the observations up to time $t_{k+1}$, i.e., $\boldsymbol{z}^{(1)},\dots,\boldsymbol{z}^{(k+1)}$. When  we utilize observations to estimate the state of the system, we say that the data are assimilated into the model, and use the notation $\widehat{\boldsymbol{x}}^{(k+1)}$ to denote an analyzed state estimate of the system at time $t_{k+1}$. When all the observations before (but not including) time $t_{k+1}$ are applied for estimating the state of the system, then we call it the forecast estimate and denote it as $\boldsymbol{x}^{(k+1)}_f$. The ensemble Kalman filter (EnKF) \cite{burgers1998analysis} follows the Monte Carlo estimation method to approximate the covariance matrix in the Kalman filter equations \cite{kalman1960new}.  Instead of modeling the exact evolution of a probability density function under nonlinear dynamics, ensemble methods maintain an empirical approximation to the target distribution in the form of a set of ensemble members $\boldsymbol{\widehat{X}}^{(k)}(i)$ for $i=1 \dots N$. We begin by initializing the state of the system for different ensemble members $\boldsymbol{\widehat{X}}^{(0)}(i)$ drawn from the distribution ${\cal{N}} (\boldsymbol{\widehat{x}}^{(0)},\boldsymbol{P}^{(0)})$, where $\boldsymbol{\widehat{x}}^{(0)}$ represents the best-known state estimate at time $t_0$, and $\boldsymbol{P}^{(0)}$ is the initial covariance error matrix.

The propagation of the state for each ensemble member over the time interval $[t_k,t_{k+1}]$ can be written as
\begin{align}
\boldsymbol{X}^{(k+1)}_f(i) &= \boldsymbol{M}_{t_k \rightarrow t_{k+1}}(\widehat{\boldsymbol{X}}^{(k)}(i)) + \boldsymbol{w}^{(k+1)}.
\label{eq:enkf_xk}
\end{align}
The prior state and the covariance matrix are approximated using the sample mean and error covariance matrix $\boldsymbol{P}^{(k+1)}_f$ as follows
\begin{align}
\boldsymbol{x}^{(k+1)}_f &= \frac{1}{N} \sum_{i=1}^N \boldsymbol{X}^{(k+1)}_f(i), \label{eq:enkf_xf}\\
\boldsymbol{A}^{(k+1)}_f(i) &= \boldsymbol{X}^{(k+1)}_f(i) - \boldsymbol{x}^{(k+1)}_f, \\
\boldsymbol{P}^{(k+1)}_f &= \frac{1}{N-1} \sum_{i=1}^N  \boldsymbol{A}^{(k+1)}_f(i) (\boldsymbol{A}^{(k+1)}_f(i))^{\text{T}}, \label{eq:enkf_pf}
\end{align}
where the superscript $\text{T}$ denotes the transpose, and $\boldsymbol{A}^{(k+1)}_f(i)$ is the anomalies between the forecast estimate for the $i$th ensemble and the sample mean. Once the observations get available at time $t_{k+1}$, the forecast state estimate is assimilated using the Kalman filter analysis equation as follows 
\begin{align}
    \widehat{\boldsymbol{x}}^{(k+1)} = \boldsymbol{x}^{(k+1)}_f + \boldsymbol{K}^{(k+1)}[\boldsymbol{z}^{(k+1)} - \boldsymbol{q}(\boldsymbol{x}^{(k+1)}_f)]. \label{eq:denkf_xa}
\end{align}

Unlike the EnKF algorithm, the DEnKF does not employ any perturbed observations. The Kalman gain matrix is computed using its square root version (without storing or computing $\boldsymbol{P}^{(k+1)}_f$ explicitly) as follows
\begin{multline}
    \boldsymbol{K}^{(k+1)} = \frac{\mathcal{A}^{(k+1)}_f(\boldsymbol{Q}^{(k+1)} \mathcal{A}^{(k+1)}_f)^{\text{T}}}{N-1} \bigg[\frac{(\boldsymbol{Q}^{(k+1)}\mathcal{A}^{(k+1)}_f)(\boldsymbol{Q}^{(k+1)}\mathcal{A}^{(k+1)}_f)^\text{T}}{N-1}   + \boldsymbol{R}^{(k+1)} \bigg]^{-1},
\end{multline}
where $\boldsymbol{Q} \in \mathbb{R}^{m \times n}$ is the Jacobian of the observation operator (i.e., $Q_{kl} = \frac{\partial q_k}{ \partial x_l}$), and the matrix $\mathcal{A}^{(k+1)}_f \in \mathbb{R}^{n \times N}$ is concatenated as follows
\begin{align}
\mathcal{A}^{(k+1)}_f = [\boldsymbol{A}^{(k+1)}_f(1), \boldsymbol{A}^{(k+1)}_f(2), \dots, \boldsymbol{A}^{(k+1)}_f(N)]. \label{eq:concat_A}
\end{align}
The anomalies for all ensemble members are then updated separately with half the Kalman gain as shown below 
\begin{align}
    \widehat{\boldsymbol{A}}^{(k+1)}(i) = \boldsymbol{A}^{(k+1)}_f(i) - \frac{1}{2}\boldsymbol{K}^{(k+1)} \boldsymbol{Q}^{(k+1)}\boldsymbol{A}^{(k+1)}_f(i). \label{eq:denkf_aa}
\end{align}
The state for all ensemble members is updated by adding ensemble anomalies to analysis state estimate and can be written as 
\begin{align}
    \widehat{\boldsymbol{X}}^{(k+1)}(i) = \widehat{\boldsymbol{x}}^{(k+1)} + \lambda \cdot \widehat{\boldsymbol{A}}^{(k+1)}(i), \label{eq:denkf_ea}
\end{align}
where $\lambda$ is the inflation factor to account for modeling errors. Inflation and covariance localization approaches are usually used in the EnKF framework to mitigate small number of ensembles \cite{houtekamer2005ensemble,attia2019dates,ahmed2020pyda}. The above ensembles are used as initial ensembles for the next assimilation cycle and the procedure is continued.

The DA procedure described so far corresponds to the full state vector of the system. However, in this study, the forward model is replaced with the surrogate model, and therefore, we perform the data assimilation in the latent space. The similar ideas have also been used in other studies that deals with the data assimilation for reduced-order models \cite{amendola2020data, peyron2021latent, maulik2022aieada}. Once the LSTM network is trained, it is used to forecast the future state of the POD modal coefficients in an auto-regressive manner \cite{pawar2021nonintrusive}. The evolution of the reduced-order model over the interval $[t_k,t_{k+1}]$ with the LSTM network is given as follows
\begin{equation}\label{eq:dyn_model_rom}
    \boldsymbol{\alpha}_f^{(k+1)} = \boldsymbol{F}_{t_k \rightarrow t_{k+1}}(\widehat{\boldsymbol{\alpha}}^{(k)}),
\end{equation}
where $\boldsymbol{\alpha}_f^{(k+1)}$ is the low-dimensional forecast estimate of the system, and $\widehat{\boldsymbol{\alpha}}^{(k)}$ is the low-dimensional analyzed state of the system at time $t_k$. Once the observations at time $t_{k+1}$ gets available, they need to be processed to obtain the \emph{latent} observations $\widetilde{\boldsymbol{\alpha}}^{(k+1)}$. Usually, the observations are available at very few sparse locations, i.e., $m << n$, and therefore, they need to be mapped from observation-space $\mathbb{R}^m$ to state-space $\mathbb{R}^n$ through some reconstruction technique.  Peyron et al. \cite{peyron2021latent} used the simple interpolation to learn this mapping while applying data assimilation to the indoor air quality problem. We utilize the POD reconstruction augmented with QR pivoting for learning the map from observations to full state, and its further details are provided in Section~\ref{sec:qr}. The analysis equation in the latent space can be written as follows
\begin{align}
    \widehat{\boldsymbol{\alpha}}^{(k+1)} = \boldsymbol{\alpha}^{(k+1)}_f + \widetilde{\boldsymbol{K}}^{(k+1)}[\widetilde{\boldsymbol{\alpha}}^{(k+1)} - \boldsymbol{q}(\widetilde{\boldsymbol{\alpha}}^{(k+1)}_f)]. \label{eq:denkf_xa_ls}
\end{align}
The Kalman gain matrix in Eq.\ref{eq:denkf_xa_ls} is calculated as follows
\begin{multline}
    \widetilde{\boldsymbol{K}}^{(k+1)} = \frac{\widetilde{\mathcal{A}}^{(k+1)}_f(\widetilde{\boldsymbol{Q}}^{(k+1)} \widetilde{\mathcal{A}}^{(k+1)}_f)^{\text{T}}}{N-1} \bigg[\frac{(\widetilde{\boldsymbol{Q}}^{(k+1)}\widetilde{\mathcal{A}}^{(k+1)}_f)(\widetilde{\boldsymbol{Q}}^{(k+1)}\widetilde{\mathcal{A}}^{(k+1)}_f)^\text{T}}{N-1}   + \widetilde{\boldsymbol{R}}^{(k+1)} \bigg]^{-1},
\end{multline}
where $\widetilde{\boldsymbol{Q}} \in \mathbb{R}^{N_r \times N_r}$ is the Jacobian of the \emph{latent} observation operator, and the matrix $\widetilde{\mathcal{A}}^{(k+1)}_f \in \mathbb{R}^{N_r \times N}$ is formed in latent space similar to Eq.\ref{eq:concat_A}.

\subsection{Reconstruction from Discrete Sensor Locations} \label{sec:qr}
As discussed previously, we need to reconstruct the full state of the system from limited number of discrete sensor locations. The low-rank approximation methods based on POD modes are one of the most popular method for the reconstruction task, where the sensor data is used to estimate the POD modal coefficients \cite{drmac2016new,manohar2018data}. Additionally, the QR decomposition with column pivoting is used for the near-optimal sensor placement in contrast to random sensor placement. We are interested in estimating the full state of the system given the sensor measurements $\boldsymbol{s} \in \mathbb{R}^m$ at discrete locations. For this case, we have $s_i=x_j$ for some $1 \le i \le m$ and $1 \le j \le n$. We can write this as follows   
\begin{equation}
    \boldsymbol{s} = \boldsymbol{\Theta} \boldsymbol{x}, 
\end{equation}
where $\boldsymbol{\Theta} \in \mathbb{R}^{m \times n}$ is constructed by taking $m$ rows of $n \times n$ identity matrix. Therefore, each row of the matrix $\boldsymbol{\Theta}$ will consist of all zeros except for the corresponding observation location, where it will have the value of one. We use the first $m$ POD modes obtained from singular value decomposition of the matrix $\boldsymbol{A}$ given in Eq.~\ref{eq:snapshot_A} for the reconstruction task and is denoted as $\boldsymbol{\Psi} \in \mathbb{R}^{n \times m}$ (i.e., the first $m$ columns of $\boldsymbol{U}$). The $\boldsymbol{\Psi}$ is also referred to as the tailored basis functions in the literature. The approximation of the full state is given by
\begin{equation}
    \boldsymbol{x} \approx \widetilde{\boldsymbol{x}} = \boldsymbol{\Psi} \mathbf{a}, \label{eq:reconstruction_tb}
\end{equation}
where $\mathbf{a} \in \mathbb{R}^m$. Once the sensor measurements gets available, the POD modal coefficients for the new sample can be calculated as follows
\begin{align}
    \boldsymbol{s} &\approx \boldsymbol{\Theta} \boldsymbol{\Psi} \mathbf{a},\\
    \mathbf{a} &= (\boldsymbol{\Theta} \boldsymbol{\Psi})^{-1} \boldsymbol{s}.
\end{align}
Once the POD modal coefficients for the new sample is determined, the full state of the system can be reconstructed using Eq.~\ref{eq:reconstruction_tb}. The \emph{latent} observations for data assimilation are computed by projecting the reconstructed data onto the POD basis functions $\boldsymbol{\Phi}$ as follows
\begin{equation}
    \widetilde{\boldsymbol{\alpha}} = \boldsymbol{\Phi}^T \widetilde{\boldsymbol{x}}.
\end{equation}

So far, we did not discuss the choice of the matrix $\boldsymbol{\Theta}$. The choice of the sensor locations can have a significant impact on the accuracy of the full state reconstruction. The sensor locations can be chosen either randomly or based on some heuristics or intuition of the physical problem. There are several techniques like optimal experimental design \cite{joshi2008sensor} and Bayesian criteria \cite{krause2008near} that can be used to determine the optimal sensor locations for moderately sized problems. The QR factorization is a powerful and robust data-driven method to determine the near-optimal sensor locations solely based on the data \cite{drmac2016new, manohar2018data}. In this method, we perform the QR decomposition with column pivoting of the matrix $\boldsymbol{\Psi}^T$. The QR decomposition calculates a column permutation matrix $\mathbfcal{C}^T \in \mathbb{R}^{n \times n}$, an orthogonal matrix $\mathbfcal{Q} \in \mathbb{R}^{m \times m}$, and an upper triangular matrix $\mathbfcal{R} \in \mathbb{R}^{m \times n}$ such that $\boldsymbol{\Psi}^T \mathbfcal{C}^T = \mathbfcal{Q}\mathbfcal{R}$. The greedy approximation of the optimal sensor locations is obtained from the first $m$ rows of matrix $\mathbfcal{C}$, i.e., by setting $\boldsymbol{\Theta} = \overline{\mathbfcal{C}}$, where $\overline{\mathbfcal{C}}$ corresponds to first $m$ rows of the matrix $\mathbfcal{C}$ \cite{manohar2018data,williams2022data}. We note here that one can also use the QR decomposition method for oversampled case, where the number of sensor exceeds the number of tailored basis functions used for reconstruction.  

\section{Results and Discussion} \label{sec:results}
We first describe the NOAA Optimum Interpolation (OI) Sea Surface Temperature (SST) V2 dataset used in this study for building a surrogate model and then integrating it within the data assimilation cycle. Then we present our numerical experimental results where we analyze the performance of our surrogate model and surrogate model assisted data assimilation framework.  

\subsection{Dataset and Preprocessing}
We use the NOAA OI SST V2 analysis dataset for building our surrogate model. The analysis uses in situ  (ship and buoy) and satellite SSTs plus SSTs simulated by the sea-ice cover. Before computing the analysis, the satellite data is adjusted for biases using the method of  Reynolds \cite{reynolds1988real} and Reynolds and Marsico \cite{reynolds1993improved}. This dataset consists of the weekly average sea surface temperature on a $1^\circ$ latitude $\times$ $1^\circ$ longitude global grid ($180 \times 360$). The SST dataset exhibits a strong periodic structure due to seasonal fluctuations. Despite this seasonal periodicity, complex ocean dynamics lead to rich flow physics in this dataset. This dataset has been used in number of recent studies on flow reconstruction \cite{callaham2019robust}, geophysical emulation \cite{maulik2020recurrent}, and dynamic mode decomposition \cite{kutz2016multiresolution}. Here, we use the data from October 1981 to  December 2000 (1000 snapshots) for building a surrogate model and the data from January 2001 to June 2018 (914 snapshots) for comparing the performance of the surrogate model for forecasting.

\subsection{Numerical Experiments}
\begin{figure*}[htbp]
\centering
\includegraphics[width=0.99\textwidth]{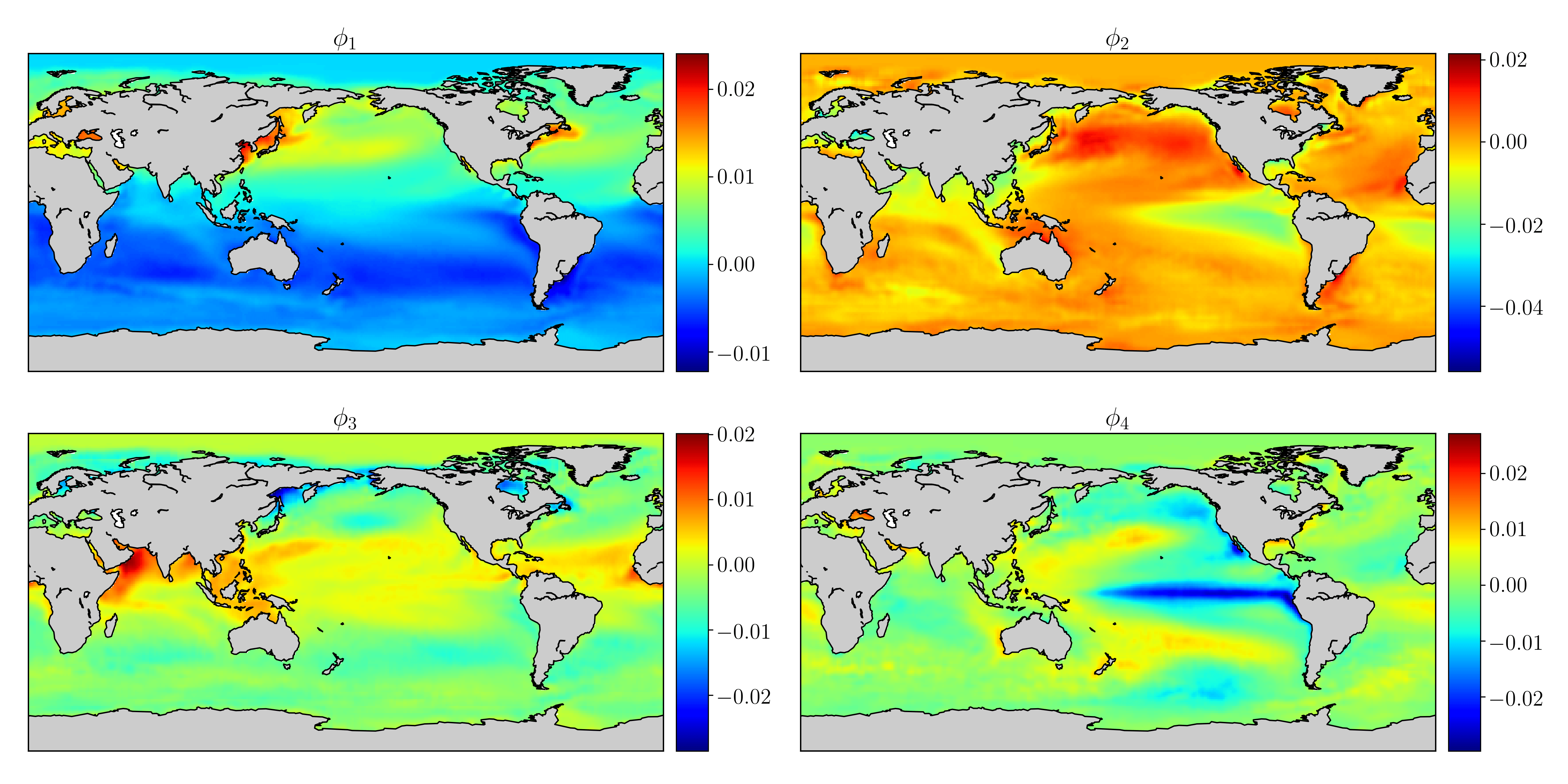}
\caption{The first four leading POD modes extracted from the SST dataset. The temporal mean was subtracted from the data and the POD modes are computed for he unsteady component of the SST field. }
\label{fig:bases}
\end{figure*}

\begin{figure*}[htbp]
\centering
\includegraphics[width=0.90\textwidth]{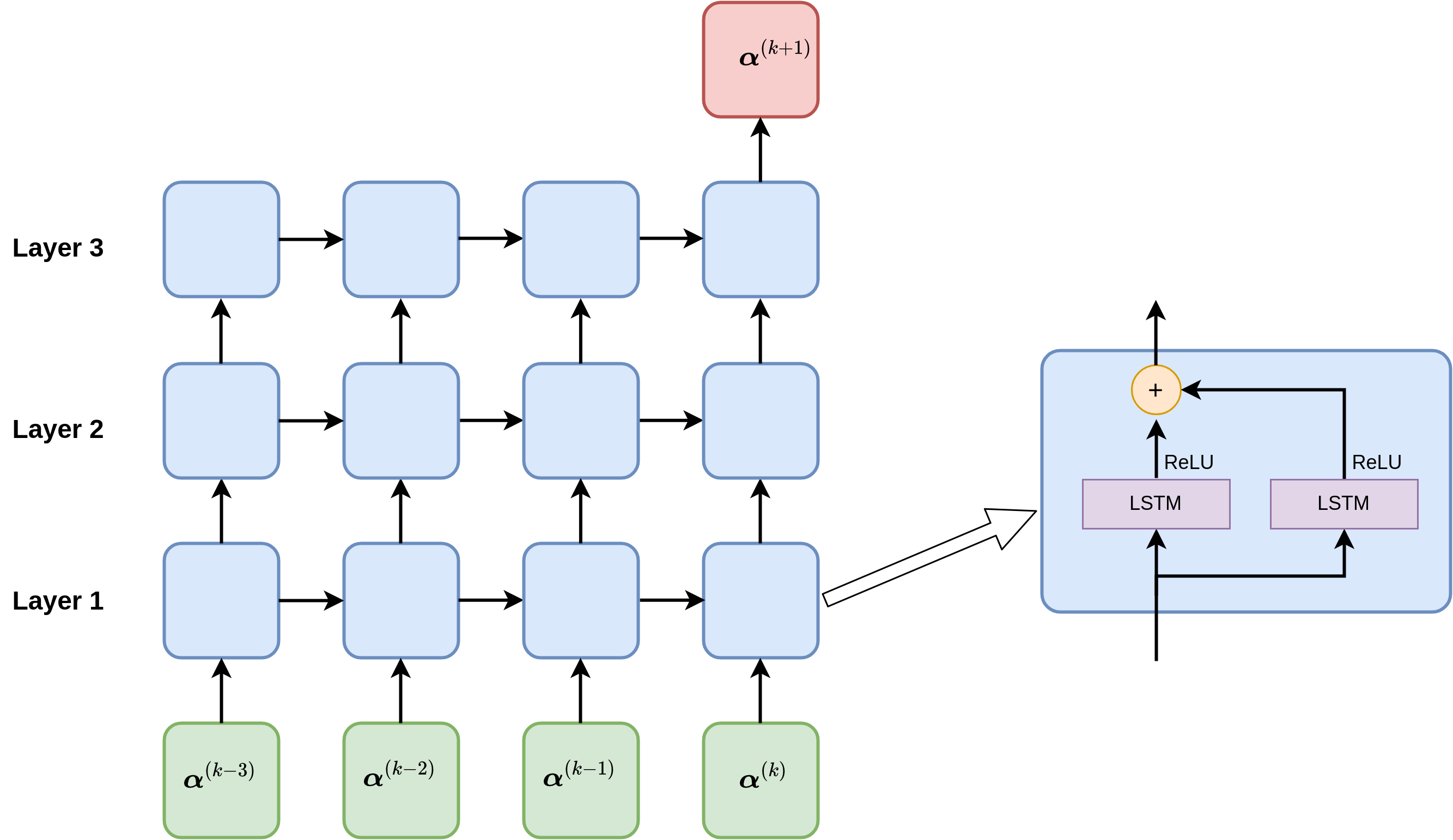}
\caption{LSTM network architectures used for modeling the dynamics of reduced-order dynamical system. The LSTM is trained to predict the future forecast state of the system based on the previous four consecutive states of the system. }
\label{fig:lstm}
\end{figure*}

In this subsection, we detail the numerical experiment design to assess the performance of NIROM and the integration of NIROM with DA, i.e., NIROM-DA. First, we use the data from October 1981 to  December 2000 (1000 snapshots) to identify the POD basis functions. The masking operation is used to remove the data that correspond to the land area prior to its utilization for surrogate modeling. The temporal mean is also subtracted from the data and the POD is carried out for the unsteady component of the SST field. The first four POD modes capture approximately 90\% of the variance of the data and are enough to capture the long-term and seasonal trends in the SST. Therefore, we fix $N_r=4$ in this study. The first four POD bases computed with POD are shown in Fig.~\ref{fig:bases}. The next step is to train the LSTM network to learn the dynamics of a reduced-order system and a lookback window of $l=4$ is used. Once the LSTM is trained, the model is deployed in auto-regressive manner for dynamical system forecasting. In the auto-regressive deployment, the initial condition for $l$ time steps is used to predict the forecast state of the system at $(l+1)$th time step. Then the state of the system from $(2)-(l+1)$ is used to determine the forecast state at $(l+2)$th time step. The LSTM network architecture used in this study employs skip-connection after every hidden layer and is displayed in Fig.~\ref{fig:lstm}. The input and output data for training the LSTM network is scaled between $[-1,1]$ to accelerate the training. The LSTM network utilizes three layers of stacked cells, 80 neurons per LSTM cell, and a ReLU activation function. The LSTM network is trained using an Adam optimizer with the learning rate of $1 \times 10^{-3}$ and the weights of the network are initialized with the random normal initializer. The hyperparameters of the network are determined based on previous studies \cite{pawar2022hyperparameter}, and considerations of computational efficiency. 

\begin{figure*}[htbp]
\centering
\includegraphics[width=0.99\textwidth]{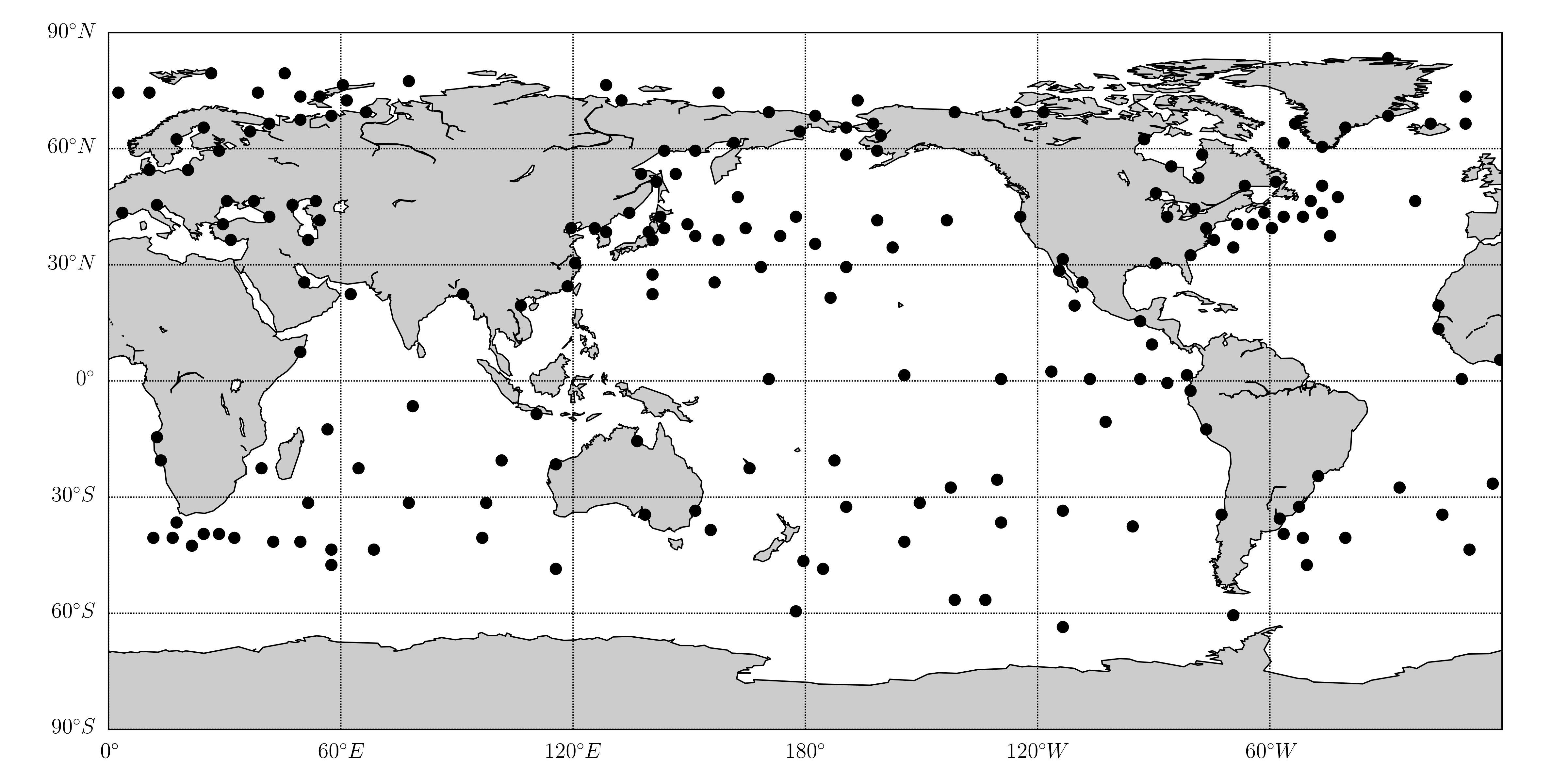}
\caption{Optimal sensor locations for reconstruction obtained using QR. These locations are informative about the ocean dynamics in contrast to random selection of sensor locations.}
\label{fig:sensors}
\end{figure*}

Once the surrogate model for SST is built, it is integrated into the DA cycle. The performance of the NIROM-DA framework is analyzed using a twin experiment for the period of January 2001 to June 2018. This data was not utilized in any stage of the NIROM construction. This ensures that there is no overlap in the training and online deployment of the surrogate model. For our twin experiment, the observations are generated by adding noise drawn from the Gaussian distribution with zero mean and the covariance matrix $\boldsymbol{R}_k$, i.e., $\boldsymbol{v}_k \sim {\cal{N}}(0,\boldsymbol{R}_k)$. We use $\boldsymbol{R}_k = \sigma_b^2 \boldsymbol{I}$, where $\sigma_b$ is the standard deviation of measurement noise and is set at $\sigma_b=1$. The observations are assumed to be collected only at discrete locations and the sensor locations are determined using the QR pivoting. We utilize only 300 discrete sensors for the reconstruction and this corresponds to less than 1\% of the full state of the system. This number is also very close to the optimal rank truncation threshold determined for this dataset from previous studies \cite{manohar2018data}. The number of tailored basis functions is also set equal to the number of sensors to avoid the intractable computation associated with the oversampled case. As discussed in Section~\ref{sec:qr}, the tailored basis functions are just the POD modes, and we can measure the variance of the data captured by these modes using their singular values. The first 300 POD modes capture approximately more than 99.5\% of the variance of the data and this leads to the improved reconstruction of small-scale features. Fig.~\ref{fig:sensors} depicts the near-optimal sensor locations determined through QR pivoting.

The initialization of ensembles is very important for the sequential DA \cite{houtekamer2016review}. We experimented with three methods for initialization of ensembles, the first method is using random snapshots collected over some time to initialize the ensembles, the second one is adding random perturbation to the full state and then projecting it onto POD basis functions, and the third method is to add a random perturbation to the reduced-order state of the system. The first method is not suitable for our problem because of the seasonality in the dataset and this leads to the phase difference between the initial condition of ensembles. The second method did not lead to sufficient variability in the initial condition of the ensembles. The third approach ensures that the initial condition for all ensembles is sufficiently different without having any phase difference. Since we are doing both the forecast and data assimilation in latent space, the approach of adding a random perturbation to the true latent space is more suitable for our problem. Once all the ensembles are initialized, the forecasting is started using the NIROM in an auto-regressive manner. The observations are assimilated every third week to correct the initial condition for the future forecast. As discussed previously, the latent observations are obtained by first reconstructing the full state using the information at discrete sensor locations and tailored basis functions. The full state is then projected onto POD modes to compute latent observations for assimilation. The number of ensembles is set at 40 and the inflation factor of 1.5 is used to account for model error.

\begin{figure*}[htbp]
\centering
\includegraphics[width=0.99\textwidth]{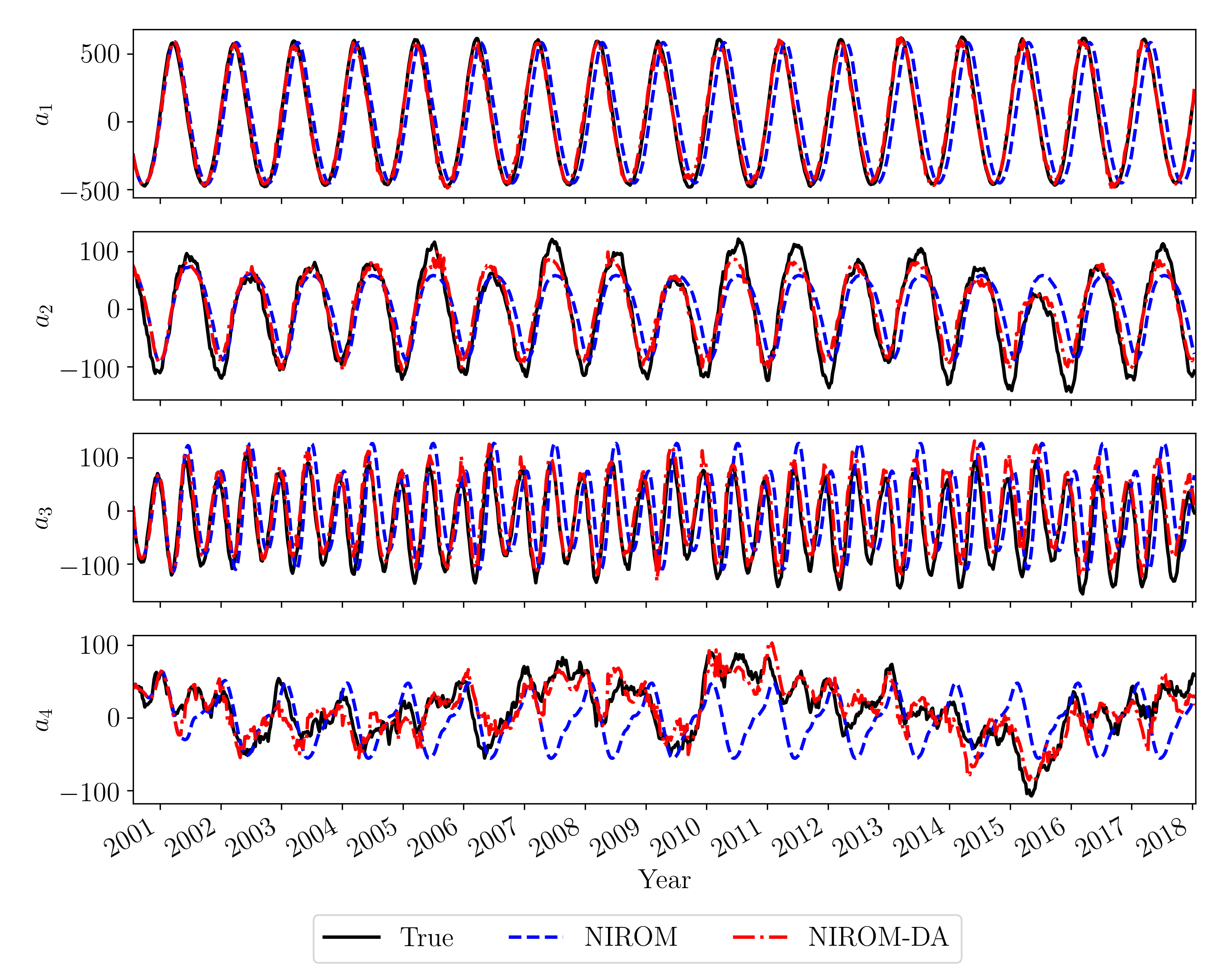}
\caption{Evolution of the POD modal coefficients for the forecasting (i.e., 2001-2018 period) of the sea surface temperature data. Predictions are started with an ensemble of 40 noisy observations.}
\label{fig:modes}
\end{figure*}

Fig.~\ref{fig:modes} shows the evolution of POD modal coefficients for the forecasting period, i.e., 2001-2018. The POD modal coefficients can be interpreted as the contribution of different spatial frequencies (i.e., basis functions) in the evolving flow. One can observe that the first three modal coefficients are responsible for capturing the seasonal dynamics, and hence do not exhibit strong chaotic behavior. The fourth modal coefficient is significantly more chaotic than the first three modal coefficients due to the stochastic nature of small-scale fluctuations. The prediction from the NIROM is quite accurate in the initial period and the error gradually increases with time. This is a well-known limitation of the auto-regressive deployment of the LSTM for modeling the evolution of dynamical systems \cite{sangiorgio2020robustness, maulik2020recurrent} and can be attributed to the error accumulation over time. The difference is particularly considerable for the first modal coefficient where there is a large phase difference between the true dynamics and predicted dynamics near the final time. Similar errors are also observed for the second and third modal coefficients along with a very inaccurate prediction for the fourth modal coefficient. The NIROM-DA framework can provide an analysis state that is very close to the true modal coefficients. The phase difference between the true and analyzed modal coefficients is very small leading to accurate modeling of seasonal dynamics. The analysis state of the fourth modal coefficient is also very close to the true state, meaning that the small-scale fluctuations are captured accurately with the NIROM-DA framework.  

\begin{figure*}[htbp]
\centering
\includegraphics[width=0.99\textwidth]{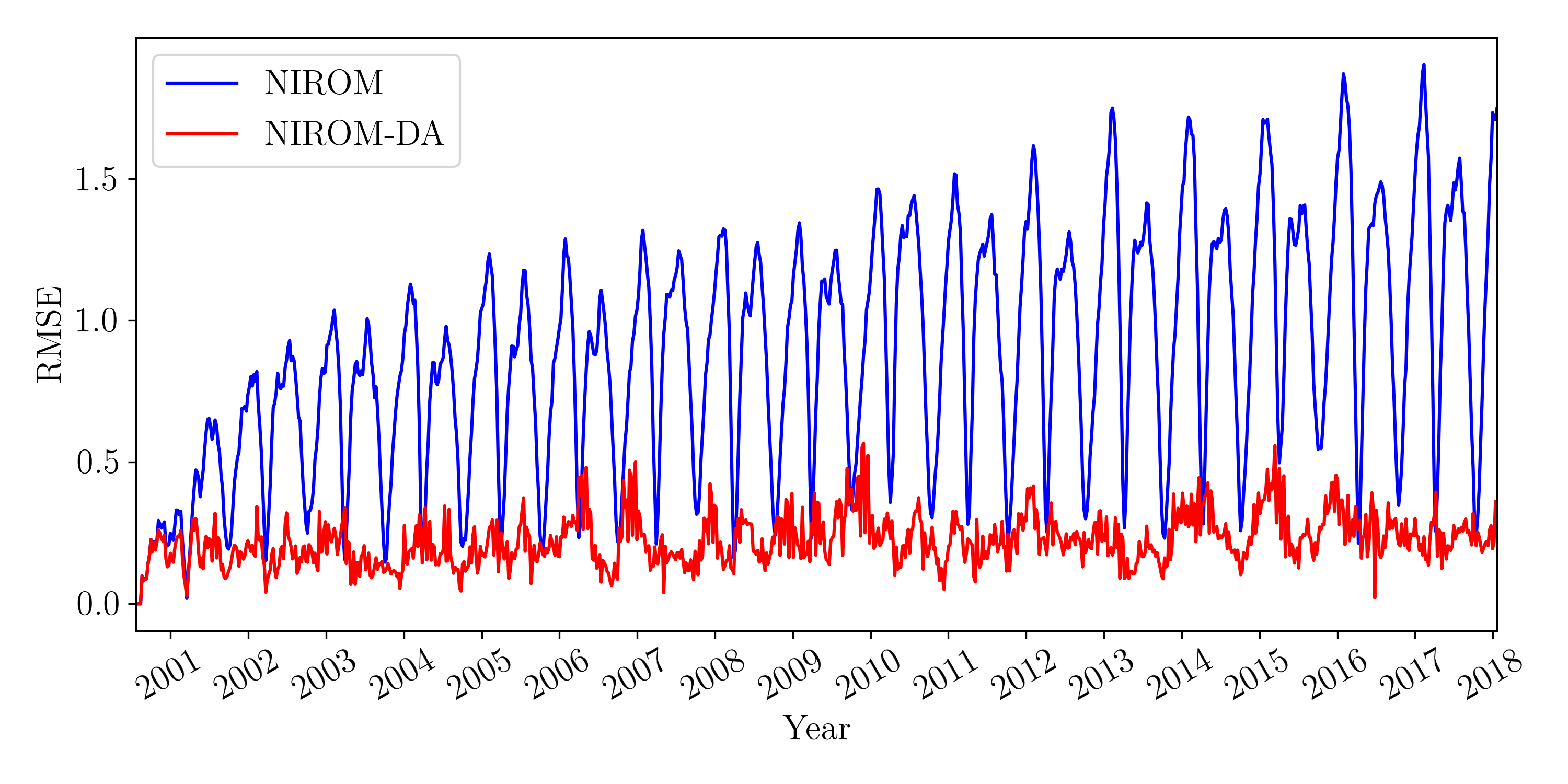}
\caption{The RMSE (in degree Celsius) for the forecast (i.e., 2001-2018 period) between the projection of observed weekly average SST data onto four POD modes and the forecast obtained from NIROM, and NIROM-DA approaches. }
\label{fig:rmse}
\end{figure*}

\begin{figure*}[htbp]
\centering
\includegraphics[width=0.99\textwidth]{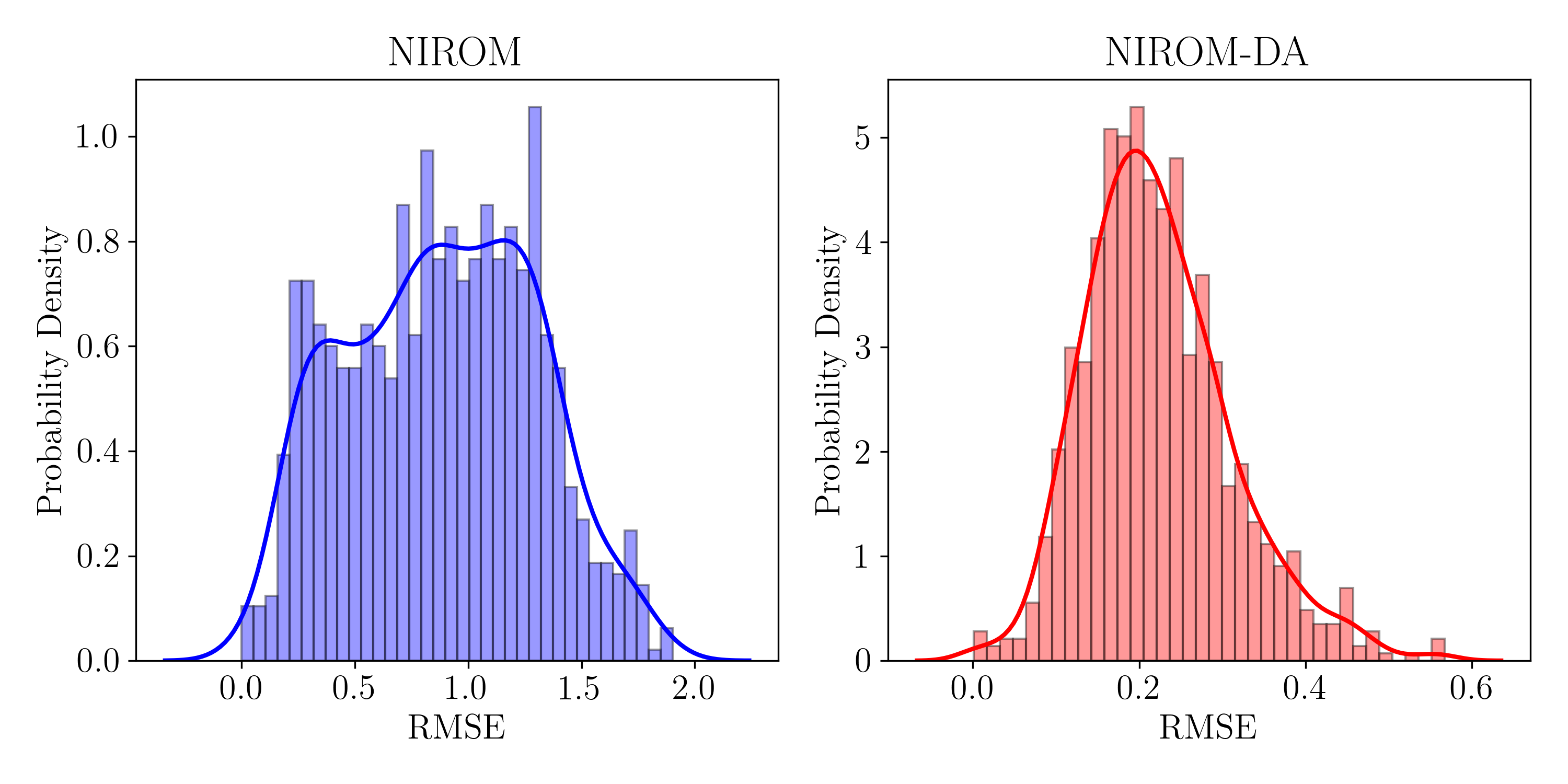}
\caption{The distribution of the forecast RMSE in degree Celsius for NIROM (left) and NIROM-DA (right).}
\label{fig:pdf_rmse}
\end{figure*}

Next, we analyze the performance of NIROM and NIROM-DA frameworks in the reconstruction of the temperature field. The temperature field is reconstructed using the predicted modal coefficients and the POD basis functions with Eq.~\ref{eq:reconstruction}. We note here that the performance of our surrogate model strategy is limited by the number of POD modes used and the energy captured by those POD modes. Thus, we can at the most recover the true projection (TP) of the analysis temperature field (also referred to as the FOM) onto the POD modes. One can also interpret TP as the filtered version of the analysis temperature field. Thus, for our analysis, the root mean squared error (RMSE) is computed between the TP temperature field and the reconstructed temperature field from NIROM and NIROM-DA frameworks. Fig.~\ref{fig:rmse} displays the evolution of RMSE in degrees Celsius for the forecasting period. Overall the RMSE for NIROM-DA is one order of magnitude less than using only NIROM for the prediction. The probability density of the RMSE for NIROM and NIROM-DA approaches is displayed in Fig.~\ref{fig:pdf_rmse}. The RMSE for NIROM lies mostly between $0.5^\circ$ Celsius to $1.5^\circ$ Celsius, while the RMSE for NIROM-DA is mostly centered around $0.2^\circ$ Celsius. This shows that a pure data-driven surrogate model can be built by exploiting the observational data collected over many decades and it can provide a sufficient level of accuracy in forecasting. Furthermore, the prediction from the forecast model can be improved using online sensor measurement in a computationally efficient manner using latent assimilation.

\begin{figure*}[htbp]
\centering
\includegraphics[width=0.99\textwidth]{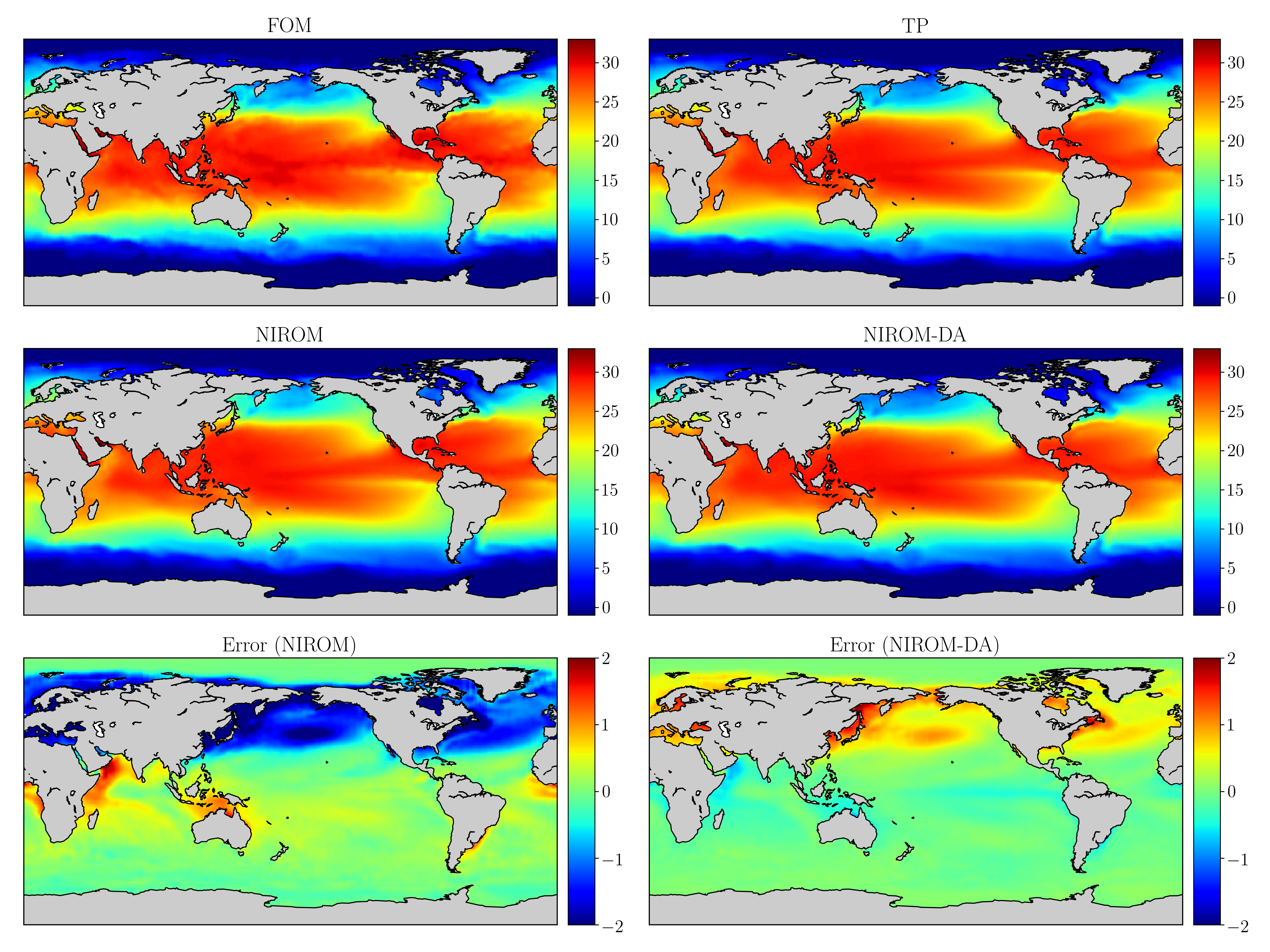}
\caption{Sample averaged temperature forecasts in degrees Celsius for the week of September 14, 2009. The FOM corresponds to actual observed averaged sea surface temperature and the TP corresponds to the projection of the FOM data onto four POD modes. The error for NIROM and NIROM-DA are calculated as the difference between the TP field and the predicted field.}
\label{fig:field_mid}
\end{figure*}

\begin{figure*}[htbp]
\centering
\includegraphics[width=0.99\textwidth]{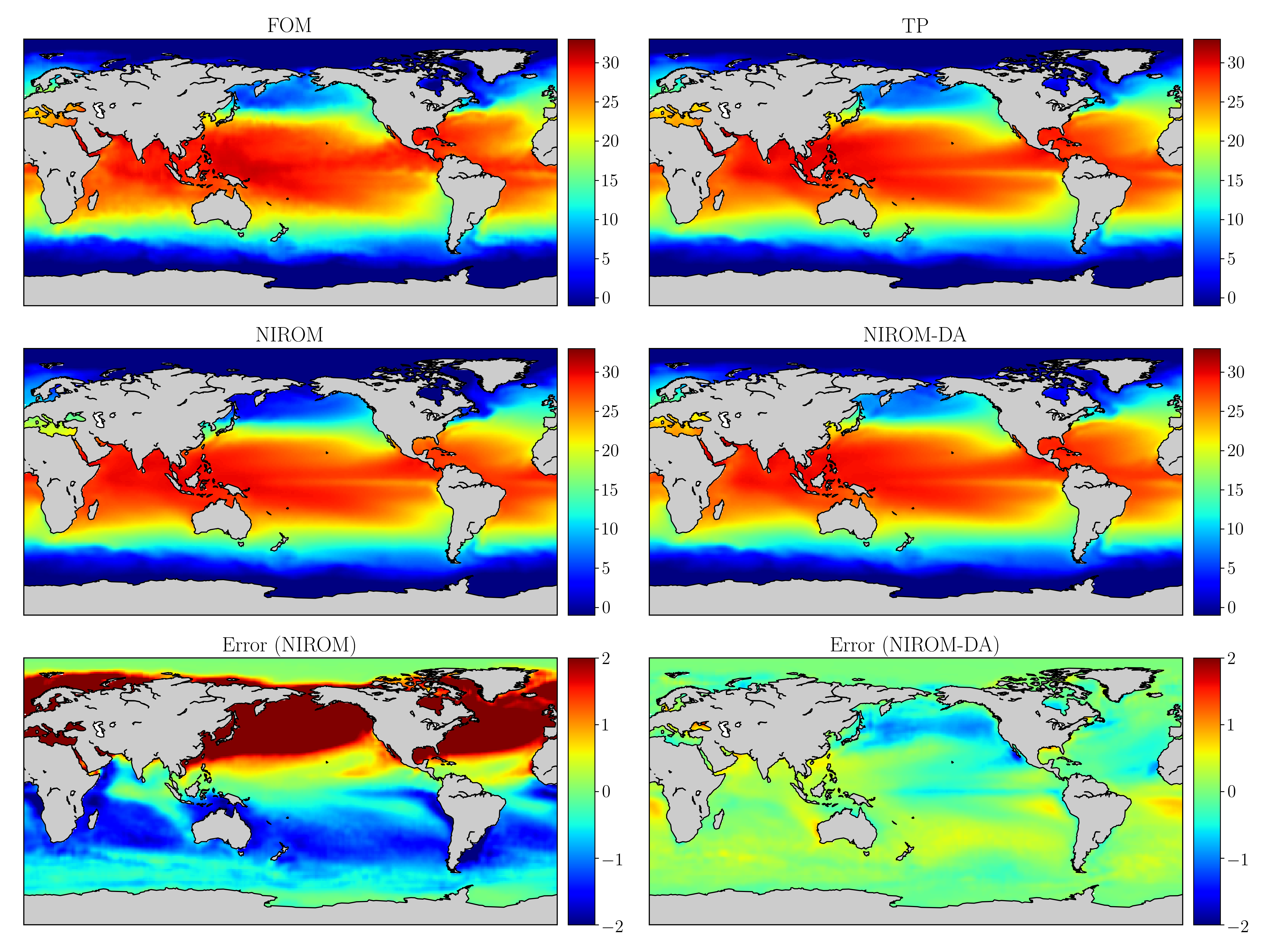}
\caption{Sample averaged temperature forecasts in degrees Celsius for the week of June 21, 2018. The FOM corresponds to actual observed averaged sea surface temperature and the TP corresponds to the projection of the FOM data onto four POD modes. The error for NIROM and NIROM-DA are calculated as the difference between the TP field and the predicted field.}
\label{fig:field_final}
\end{figure*}

The quantitative findings for the NIROM and NIROM-DA frameworks are further supported by the visualization of the temperature field at two different times. Fig.~\ref{fig:field_mid} displays the temperature field on September 14, 2009 and the temperature field on June 21, 2018 is shown in Fig.~\ref{fig:field_final}. As we are retaining only four POD modes for the surrogate model, some of the small-scale features of the exact temperature field (i.e., FOM) are not captured by TP. While these small-scale features can be captured by retaining a large number of POD modes, training a recurrent neural network to learn the dynamics of lower-energy POD modes will require additional treatment. One potential solution for this can be the non-linear proper orthogonal decomposition (NLPOD) where an autoencoder is used to compress a large number of POD modal coefficients \cite{ahmed2021nonlinear}. From Fig.~\ref{fig:field_mid}, it is seen that even though larger structures in the flow are captured by NIROM, the error is higher for NIROM compared to NIROM-DA, especially in the northern hemisphere region. Similar observations are also noted in Fig.~\ref{fig:field_final}, where the error for NIROM is larger than the NIROM-DA in both the northern and southern hemisphere regions.  

\section{Concluding Remarks} \label{sec:conlusion}
In this paper, we propose a novel framework to construct a non-intrusive reduced order model (NIROM) for geophysical flow emulation and integrate this low-dimensional model within the deterministic ensemble Kalman filter (DEnKF) algorithm to assimilate sparse and noisy observations in the latent space. The NIROM is based on a proper-orthogonal decomposition to identify the dominant modes from the data and then uses a recurrent neural network to learn the dynamics of low-dimensional latent space. This surrogate modeling strategy exploits the archival of observational data without relying on any kind of governing equations. One of the critical components of the proposed framework is the reconstruction of full state form sparse and noisy discrete sensor measurement. This is achieved using the least square estimation to map the information at near-optimal sensor locations determined through QR pivoting to the full state of the system. The latent observations for assimilation are generated by projecting this full state onto the POD modes. 

We demonstrate the performance of our framework for the NOAA Optimum Interpolation Sea Surface Temperature (SST) analysis dataset. The NIROM is able to achieve a stable long-range forecast along with predicting larger structures in the temperature field with a sufficient level of accuracy. Once the NIROM is integrated within the DEnKF, the prediction is improved quantitatively by almost one order of magnitude. The results show that the NIROM can be readily coupled with the DA and latent space trajectory can be corrected every time sparse and noisy observations get available. With this framework, the computational saving is achieved through the replacement of a forward numerical solver with a data-driven model, and assimilation in latent space in contrast to the full state of the system. The proposed framework is extremely flexible and other algorithms can be easily accommodated. For example, one can use algorithms like convolutional autoencoder for dimensionality reduction, or shallow decoder for the reconstruction of full state from sparse discrete measurements. Furthermore, the NIROM can be easily differentiated with automatic differentiation and can be utilized in variational DA settings.        

Finally, we point out that there is a number of future research directions that one can take. One such future direction is to improve the performance of the surrogate model by incorporating more modes to capture the low-energy content and frameworks like non-linear proper orthogonal decomposition \cite{ahmed2021nonlinear} can be exploited for that. Another future research direction is to enhance the robustness of a long short-term memory network by reducing error accumulation during its auto-regressive deployment \cite{sangiorgio2020robustness}. Our future work will also include exploring multi-fidelity data assimilation where a few ensembles of high-fidelity numerical solvers will be complemented with a large ensemble of data-driven models for forecasting \cite{popov2021multifidelity} and employ a shallow decoder for full state reconstruction \cite{erichson2020shallow}.

\section*{Acknowledgements}
This material is based upon work supported by the U.S. Department of Energy, Office of Science, Office of Advanced Scientific Computing Research under Award Number DE-SC0019290. O.S. gratefully acknowledges their support. 
Disclaimer: This report was prepared as an account of work sponsored by an agency of the United States Government. Neither the United States Government nor any agency thereof, nor any of their employees, makes any warranty, express or implied, or assumes any legal liability or responsibility for the accuracy, completeness, or usefulness of any information, apparatus, product, or process disclosed, or represents that its use would not infringe privately owned rights. Reference herein to any specific commercial product, process, or service by trade name, trademark, manufacturer, or otherwise does not necessarily constitute or imply its endorsement, recommendation, or favoring by the United States Government or any agency thereof. The views and opinions of authors expressed herein do not necessarily state or reflect those of the United States Government or any agency thereof.

\appendix


\bibliographystyle{unsrt} 
\bibliography{ref}

\end{document}